\documentclass[conference]{IEEEtran}
\ifCLASSINFOpdf
\else
\fi
\usepackage[english]{babel}
\usepackage[T1]{fontenc}

\usepackage[utf8x]{inputenc}
\usepackage{cite}
\usepackage{graphicx}
\usepackage{color}
\usepackage{amssymb}
\usepackage{amsthm}
\usepackage{amsxtra}
\usepackage{amsmath}
\usepackage{multirow}
\usepackage{epsfig}
\usepackage{comment}
\usepackage{subfigure}
\usepackage{psfrag}
\usepackage{url}
\usepackage{textcomp}
\usepackage{enumerate}
\usepackage[printonlyused]{acronym}
\usepackage{balance}
\usepackage{algorithm}
\usepackage{algpseudocode}
\usepackage{nomencl}
\usepackage{units}
\usepackage{pstricks, pst-plot, pst-grad, pst-node, pstricks-add}
\usepackage{pifont}
\usepackage{glossaries}
\usepackage{siunitx}

\newrgbcolor{darkgreen}{0.2 0.6 0.2}
\newrgbcolor{darkred}{0.6 0.2 0.2}
\newrgbcolor{darkblue}{0.2 0.2 0.6}
\newrgbcolor{purple}{0.2 0.1 0.9}
\begin{document}

\makeatletter
\def\ps@IEEEtitlepagestyle{
  \def\@oddfoot{\mycopyrightnotice}
  \def\@evenfoot{}
}
\def\mycopyrightnotice{
  {\footnotesize
  \begin{minipage}{\textwidth}
  \centering
 Copyright~\copyright~2019 ITG VDE SCC. Personal use of this material is permitted. Permission from ITG/VDE must be obtained for all other uses, in any current or future media, including reprinting/republishing this material for advertising or promotional purposes, creating new collective works, for resale or redistribution to servers or lists, or reuse of any copyrighted component of this work in other works.
  \end{minipage}
  }
}
\title{Cloud Control AGV over Rayleigh Fading Channel - The Faster The Better}



%

\author{
\IEEEauthorblockN{Shreya Tayade\IEEEauthorrefmark{1},
Peter Rost\IEEEauthorrefmark{2},
Andreas Maeder\IEEEauthorrefmark{2} and 
Hans D. Schotten\IEEEauthorrefmark{1}}
\IEEEauthorblockA{\IEEEauthorrefmark{1}Intelligent Networks Research Group, German Research Center for Artificial Intelligence, Kaiserslautern, Germany\\
Email: \{Shreya.Tayade, Hans\_Dieter.Schotten\} @dfki.de}
\IEEEauthorblockA{\IEEEauthorrefmark{2}Nokia Bell Labs,
Munich, Germany\\
Email: \{peter.m.rost, andreas.maeder\}@nokia-bell-labs.com}
}

\maketitle

\begin{abstract}
This paper analyzes the stability of the control system of an Autonomous Guided Vehicle (AGV) using a central controller. The control commands are transmitted to an AGV over a Rayleigh fading channel causing potential packet drops. This paper analyzes the mutual dependencies of control system and mobile communication system. Among the important parameters considered are the sampling time of the discrete control system, the maximum tolerable outages for the control system, the AGV velocity, the number of users, as well as mobile communication channel conditions. It is shown that increasing the velocity of an AGV leads to a lower risk of instability due to the higher time-variance of the mobile channel. While this still is a 'sandbox' example, it shows the potential for a manifold co-optimization of control systems operated over imperfect mobile communication channels.
\end{abstract}


%
\IEEEpeerreviewmaketitle
\section{Introduction} 
In industrial automation, coordination and cooperative operation of mobile robots have gained significant importance. The recently launched HeathrowPods at London Heathrow Airport are a good example of centralized control for multiple Autonomous Guided Vehicles (AGVs). Many new use-cases require joint operation of multiple robots that are controlled from within the edge cloud. 
A decently large body of research is already existing on simultaneous coordination and synchronization of mobile robots \cite{aguirre2011remote,Kanjanawanishkul,multiplerobots,olmi2008coordination}. 
There are a number of challenges when controlling robots through a centralized cloud controller, since control commands are sent over a  bandwidth limited, time varying wireless channel. The impact of the limited channel and cloud resources on the latency constraint applications is analyzed in \cite{tayadeICC, tayade2017device}. Furthermore, one of the challenges is to deal with unreliable communication between the cloud controller and the robot. A failure or a delay in the network can cause instabilities of the robot control. Therefore, it is crucial to analyze the performance of the control system over an unreliable communication network.

The influence of the communication network on the control system is studied rigorously in the past decade \cite{4118476foundation,1310461comm_constraints,1310480NoisyChannel,1661825anytime,1333206stochasticcontrol}.  
The authors in \cite{1310461comm_constraints} evaluate the minimum data rate required, between the plant and the controller to maintain a stable and observable LTI control system. It shows that to have a stable LTI system, the rate should be greater than the sum of all logarithmic eigenvalues of its system matrix. In \cite{1310480NoisyChannel}, the stability criterion is determined in terms of distortion measure generated by the source encoder and decoder. The stability and observability is analyzed for a noiseless digital channel, delayed communication channel, erasure channel and memory-less Gaussian channel with limited power constraints. The paper, \cite{1333206stochasticcontrol}, analyzes the impact of communication channel on the Linear Quadratic Gaussian (LQG) control problem. 
It reveals that the information pattern between the encoder and decoder has a significant impact on the stability of the control system. The author in \cite{4118476foundation}, designs an optimal controller and state estimator for a LTI system, over a TCP and an UDP implemented communication network. It evaluates the threshold error probabilities to have a stable control system. The effect of network delays on stability of the general linear and non-linear control system has been studied in \cite{Zhang:2001:SAN:934787,gracia, nof2009springer, Astrom:1990:CST:78995, MIT}. 

The papers \cite{4118476foundation,1310461comm_constraints,1310480NoisyChannel,1661825anytime,1333206stochasticcontrol, Zhang:2001:SAN:934787,gracia, nof2009springer}, present the effects of communication channel on a generic linear control system. However, most of the practical use-cases involve a non-linear, time varying, in-homogeneous control system. Therefore, in this paper a stability performance of a more practical control system of an AGV, in  presence of a wireless communication channel is investigated. Moreover, the objective of the papers in \cite{1310461comm_constraints,1310480NoisyChannel,1661825anytime,1333206stochasticcontrol}, is to design an optimal controller, encoders and decoders to retain the stability of a control system. In this paper, we investigate the fading effects on the stability performance of an AGV control system. 
The authors in \cite{aguirre2011remote} propose a control law that would sustain network delays so that the AGV remains in a stable state. The control law proposed uses a cascaded control system that predicts the position of an AGV as long as the delay exists in the network. On the contrary, this paper evaluates an upper bound on the consecutive channel outage to retain the stability of an AGV control system. The objective is to exploit the channel correlation property to optimize the communication network while simultaneously maintaining the control system stability. Furthermore, the repercussions of the communication channel in the control system were studied for a stationary controller and an actuator. We analyzed the stability performance of an mobile AGV actuator over a Rayleigh fading channel.  This paper provides insights on the relation between the AGV velocity, the channel outages and the control system stability.





In Section \ref{sec: system:model}, the control system of a central-controlled AGV over a fading channel is presented. In Section \ref{sec:delay tolerance}, the optimization problem is designed to evaluate the maximum communication outages that a stable control system can tolerate. Section~\ref{sec:fadingChannel} describes the fading channel and its error probabilities. Mutual dependencies between the control and communication and the results are discussed in Section~\ref{sec:results}. Finally, conclusions are discussed in Section~\ref{sec:conclusion}.

\section{System Model} \label{sec: system:model}
The system consists of an edge cloud that controls $N$ AGVs. The edge cloud sends the control commands to the AGVs every $T_s$\,seconds. The control commands are sent over a wireless channel of bandwidth $B$. The bandwidth is assumed to be equally shared among the $N$ AGVs. The control commands sent to a single AGV $i\in\{1, 2, \dots N\}$ are encoded with $D_i$ data bits. We use $\gamma_i$ to denote the long-term signal-to-noise ratio (SNR) for AGV $i$. 

The AGV need to trace the complete pre-defined reference track $X_r(t)$ = $[x_r(t); y_r(t); \theta_r(t)]^\mathsf{T}, t\in[0; T]$. The reference track is described by coordinates $x_r(t)$ and $y_r(t)$, respectively, and an angle $\theta_r(t)$, which gives the orientation of an AGV with respect to the X-axis. The control input sent from the controller to the $i^\text{th}$ AGV consist of the intended translational velocity $\nu(t)$ and rotational velocity $\omega(t)$. The reference track and the control input are distinct for each AGV. As the control input is sent every $T_s$ seconds, $T_s$ also represents the sampling time period. Let $k$ denote the time instance $t_k = k\cdot T_s$. The sampled reference track $X_r(k)=X_r(t_k)$ for each AGV is known to the central controller for every time sample $t_k$. 

\begin{figure*}
    \centering
  \includegraphics[scale = 0.5]{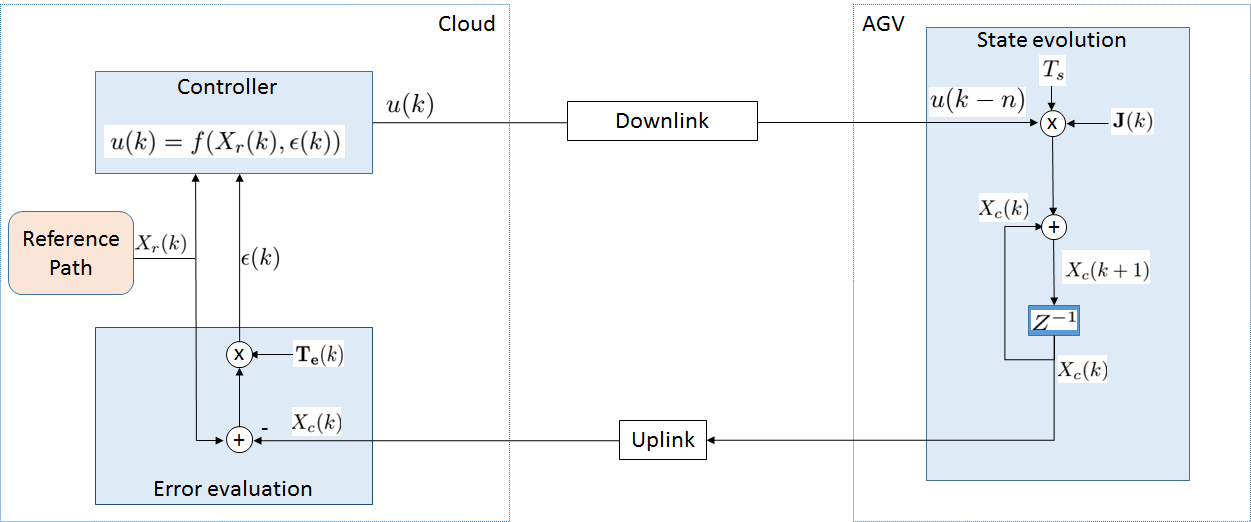}
    \caption{System model}
    \label{fig:ch3:1}
\end{figure*}

\subsection{Centralized control system of a single AGV} \label{sec: control system}
The control system of an AGV consists of a controller, an error evaluation and the state evolution functional block as shown in Figure \ref{fig:ch3:1}. The AGV has to follow a reference path given as $X_r(k)$ = $[x_r(k); y_r(k); \theta_r(k)]$. The  actual position of an AGV is denoted by $X_c(k)$ = $[x_c(k); y_c(k); \theta_c(k)]$. At $k = 0$, AGV position and reference position are identical, i.e.\, $X_c(0) = X_r(0)$. At later time instances, the error $\epsilon(k)$ is determined based on the difference between the reference path and the AGV position. Based on the error, the controller determines the control input $u(k)$. The control input is sent to an AGV over a downlink channel as shown in Figure \ref{fig:ch3:1}. Based on this input, the state $X_c(k)$ of the AGV evolves assuming constant $u(k)$ over a sampling time period $T_s$ (zero order hold). If $n$ downlink transmissions are lost due to communication channel outages, then the control input signal $u(k-n)$ is still applied at time instance $t_k$.
The resulting AGV position is transmitted back to the controller over the wireless channel to be used for the next control input calculation. Note that for the sake of brevity, we assume that the uplink is an ideal channel without channel outages and it is left for future investigation to quantify the impact of an imperfect uplink channel.
\paragraph{Error evaluation} The error evaluation determines the error $\epsilon(k)$ between the actual position $X_c(k)$ of an AGV and the reference position $X_r(k)$ as
\begin{eqnarray}
        \epsilon(k) &=& \left[\begin{array}{c}x_e(k)\\y_e(k)\\\theta_e(k)\end{array}\right] \nonumber \\
    &=& \left(\begin{array}{ccc}\cos\theta_c(k) & \sin\theta_c(k) & 0\\-\sin\theta_c(k) & \cos\theta_c(k) & 0\\0 & 0 & 1\end{array}\right) \left(X_r(k) - X_c(k)\right) \nonumber \\
     &=& \mathbf{T_e}(k) \left(X_r(k) - X_c(k)\right),
   \label{eq:ch3.ieee.10}
\end{eqnarray}
where $\mathbf{T_e(k)}$ is the rotational matrix, $x_e(k)$, $y_e(k)$ is the error determined as a result of difference in x and y coordinates, and $\theta_e(k)$ is the error in the orientation of an AGV. 
\paragraph{Controller} The controller calculates the control input $u(k)$ according to the error $\epsilon(k)$, determined in \eqref{eq:ch3.ieee.10}, by the control law as in \cite{Kanayama1990}: 
\begin{eqnarray}
    u(k) &=& \left[\begin{array}{c}\nu(k)\\ \omega(k)\end{array}\right] \nonumber \\
    &=& \left[\begin{array}{c} \nu_r(k) \cos\theta_e(k) + K_x x_e(k)\\ \omega_r(k) + \nu_r(k) \left[ K_y y_e(k) + K_\theta \sin\theta_e(k) \right] \end{array}\right], \label{eq:ch3.ieee.15}
\end{eqnarray}
where $K_x\,[\unit{s^{-1}}]$, $K_y\,[\unit{m^{-1}}]$ and $K_\theta\,[\unit{m^{-1}}]$ are constants which impact the convergence rate of the control system. The reference velocities $\nu_r(k)$ and $\omega_r(k)$ are evaluated from the reference path as $\nu_r(k)= \sqrt{\dot{x}_r^2(k) + \dot{y}_r^2(k)}$ and $\omega_r(k) = \dot{\theta_r(k)}$. 
\paragraph{State evolution}
The state evolution in the AGV is described by the AGV position over time, when the control input $u(k)$ is applied. The discrete time approximation of an AGV position is evaluated by solving the differential equation as stated in \cite{Kanayama1990}.

If $T_s$ is the sampling time period of a system, the difference equation for the position state evolution of an AGV is
\begin{align}
\left[\begin{array}{c} x_c(k+1)\\  y_c(k+1)\\ \theta_c(k+1)\end{array}\right] &= \left[\begin{array}{c} x_c(k)\\  y_c(k)\\ \theta_c(k)\end{array}\right] + T_s \cdot \mathbf{J}(k) \left[ \begin{array}{c} \nu(k) \\ \omega(k)\end{array} \right] \\
X_c(k+1) & =  X_c(k) +  T_s \cdot \mathbf{J}(k) \cdot u(k).\label{eq:ch3.ieee.20}
\end{align}
where $J(k)$ is given as
\begin{align}\label{eq:ch3.ieee.19}
\mathbf{J}(k) = \left(\begin{array}{cc} \cos\theta_c(k)& 0\\ \sin\theta_c(k) & 0 \\ 0 & 1\end{array}\right). 
\end{align}

\subsection{Downlink outages}
In presence of downlink channel outages, at time $t_k$, the AGV applies the old control input and attain the next position state $X_c(k+1)$. The state evolution of an AGV with $n$ consecutive channel outages is given as 
 \begin{equation} \label{eq:ch3.ieee.34}
 X_c(k+1) =  X_c(k) + T_s \cdot \mathbf{J}(k) \cdot u(k-n)
 \end{equation}
 where $u(k-n)$ is the zero-order hold control input obtained from \eqref{eq:ch3.ieee.15}:
 \begin{eqnarray}
    u(k-n) &=& \left(\begin{array}{c}\nu(k-n)\\ \omega(k-n)\end{array}\right). \label{eq:ch3.ieee.30} 
\end{eqnarray}
The parameter $n$ represents the consecutive outages in the mobile communication channel. If $n$ is large, the control inputs applied to an AGV are outdated and may cause system instabilities such that in the worst case, the AGV will not follow the reference path. Therefore, it is important to analyze the maximum outages $n_\text{max}$ that the central control system of an AGV can tolerate.

\subsection{Stability}
The AGV control system is stable if $\epsilon(k) \to 0$ as $k \to \infty$. The system is discrete, non-linear, and time varying. In order to determine the stability of the system, the non-linear system is linearized by taking the derivative with respect to the position state variables. 
The discrete position state equation is given in \eqref{eq:ch3.ieee.34}.
Substituting the values of $u(k-n)$ and $\mathbf{J}(k)$ from \eqref{eq:ch3.ieee.30} and \eqref{eq:ch3.ieee.19} in \eqref{eq:ch3.ieee.34}, it follows
\begin{eqnarray}
F1 := x_c(k+1) = x_c(k) + T_s \cdot \cos\left[\theta_c(k)\right] \cdot \nu(k-n)  \label{eq:ch3.ieee.40}\\ 
F2 := y_c(k+1) = y_c(k) + T_s \cdot \sin\left[\theta_c(k)\right] \cdot \nu(k-n)   \label{eq:ch3.ieee.41}\\
F3 := \theta_c(k+1) = \theta_c(k) + T_s \cdot\omega(k-n).   \label{eq:ch3.ieee.42}
\end{eqnarray}
The equilibrium point for an AGV is reached if the actual position and the reference position are equal, i.e.\  $X_c(k)$ = $X_r(k)$, and the error $\epsilon(k)$ converges to $0$. The stability of the system is evaluated by linearization of a non-linear control system at the equilibrium point $X_r(k) = X_c(k)$, i.\,e. at $\epsilon(k) = 0$. In order to linearize the non-linear control system in the form of 
\begin{align}\label{eq:ch3.ieee.45}
X_c(k+1) = \mathbf{A}(k) X_c(k) + \mathbf{B}(k) u(k-n),
\end{align}
the matrices $\mathbf{A}(k)$ and $\mathbf{B}(k)$ need to be determined using the Jacobian of the state space equations \eqref{eq:ch3.ieee.40} - \eqref{eq:ch3.ieee.42}. By taking derivative of $F1$, $F2$, and $F3$ with respect to $X_c(k)$, and substituting $\epsilon(k) = 0$, the linearized time varying system is obtained with
\begin{equation}\label{eq:ch3.ieee.50}
\mathbf{A}(k) = \left(\begin{array}{ccc} \frac{\partial{F1}}{\partial{x_c}} & \frac{\partial{F1}}{\partial{y_c}} & \frac{\partial{F1}}{\partial{\theta_c}} \\ 
\frac{\partial{F2}}{\partial{x_c}} & \frac{\partial{F2}}{\partial{y_c}} & \frac{\partial{F2}}{\partial{\theta_c}} \\
\frac{\partial{F3}}{\partial{x_c}} & \frac{\partial{F3}}{\partial{y_c}} & \frac{\partial{F3}}{\partial{\theta_c}}
\end{array}\right),
\end{equation} and 
\begin{eqnarray}\label{eq:ch3.ieee.51}
\mathbf{B}(k) = \left(\begin{array}{ccc} \frac{\partial{F1}}{\partial{\nu}} & \frac{\partial{F1}}{\partial{\omega}} \\ 
\frac{\partial{F2}}{\partial{\nu}} & \frac{\partial{F2}}{\partial{\omega}} \\
\frac{\partial{F3}}{\partial{\nu}} & \frac{\partial{F3}}{\partial{\omega}} 
\end{array}\right)  
= \left(\begin{array}{ccc} \cos\theta_c(k) & 0 \\ 
\sin\theta_c(k) & 0 \\
0 & 1 
\end{array}\right).
\end{eqnarray}
\begin{figure*}[t!]
\begin{equation}\label{eq:ch3.ieee.55}
\mathbf{A}(k) = \left(\begin{array}{ccc} 1- T_s \cdot K_x \cdot \cos \left[\theta_c(k) \right] \cdot \cos\left[\theta_c(k-n)\right] & - T_s\cdot K_x \cdot \cos\left[\theta_c(k)\right] \cdot \sin\left[\theta_c(k-n) \right] & - T_s \sin\left[\theta_c(k) \right] \nu(k-n) \\ 
-T_s \cdot K_x \cdot \sin\left[\theta_c(k)\right] \cdot \cos\left[\theta_c(k-n) \right] & 1-T_s\cdot K_x \cdot \sin \left[\theta_c(k)\right] \cdot \sin\left[\theta_c(k-n)\right] & T_s \cdot \cos \left[ \theta_c(k)\right] \nu (k-n) \\
T_s \cdot k_y \sin\left[\theta_c(k-n)\right] \nu(k-n) & - T_s \cdot k_y \cos\left[\theta_c(k-n)\right] \nu(k-n) & 1-T_s\cdot k_\theta \nu(k-n)
\end{array}\right)
\end{equation}
\end{figure*}

Let $\{\lambda_1(k), \dots, \lambda_M(k)\}$ = Eig($\mathbf{A(k)}$) be Eigenvalues of $\mathbf{A}(k)$. In case of a linear time invariant system (LTI), the system is stable if \cite{Astrom:1990:CST:78995}
\begin{equation}\label{eq:ch3.ieee.56}
    \forall i,k: 0 < |\lambda_i(k)| < 1.
\end{equation} To check the stability of a time varying system, the test for stability needs to be repeated for each time instance $t_k$.  

\section{Outage Tolerance and Stability}\label{sec:delay tolerance}
In order to ensure robust and stable control of an AGV, it is necessary to evaluate the maximum number of consecutive outages $n_\text{max}$ that an AGV control system can sustain without becoming unstable. The maximum consecutive outages $n_\text{max}$ is defined as the outage tolerance of a centralized AGV control system. If control input is sent more frequently to an AGV, it would increase the data rate and hence the system overhead. If the outage tolerance $n_\text{max}$ of the control system is known, it can unfold the optimum rate at which the control input should be sent to an AGV. This leads to a more efficient usage of bandwidth, reduced signalling overhead, and less computational load at the controller. 

Hence, we are interested in the maximum $n_\text{max}$ such that the stability condition $\forall i,k: 0 < |\lambda_i(k)| < 1$ is fulfilled. More formally,
\begin{align}
n_\text{max} & =   \max\limits_{\forall n\in \mathbb{Z}: \text{(\ref{eq:ch3.ieee.56}) holds}} n.
\label{eq:ch3:ieee:59}
\end{align}

\section{Fading Channel}\label{sec:fadingChannel}
The controller sends control input over a Rayleigh fading channel of bandwidth $B$ to the $N$ AGVs. For the sake of brevity, bandwidth is assumed to be equally distributed among $N$ AGVs, leaving the optimal resource allocation and scheduling for future work. The control signal transmitted to $i^{th}$ AGV, consist of $D_i$ data bits and is sent after $T_s$ seconds. The $N$ AGVs trace the reference path from different starting points with different velocities and will experience different channel gains. The error probabilities evaluated in this section correspond to the fading experienced by a single AGV. 
The required spectral efficiency $R_i$ for $i^{th}$ AGV can be written as 
\begin{eqnarray}
    R_i = \frac{D_i \cdot N}{T_s \cdot B}.
\end{eqnarray}
If $\gamma_i$ is the average SNR for the $i^\text{th}$ AGV, the minimum threshold SNR to successfully decode the received signal is given as 
\begin{equation} \label{eq:ch3.ieee.70}
   \gamma_\text{th} = \frac{(2^R -1)}{\gamma_i}.
\end{equation}
For simplicity, Shannon's capacity is considered, in future, more accurate models can be taken into consideration. Since a Rayleigh fading channel is considered, the probability of an error during the current transmission of the control inputs is given as
\begin{equation}\label{eq:ch3.ieee.71}
P_{e}(1) = 1 - \exp{\left(-\gamma_\text{th}\right)}.
\end{equation}

As the channel is time varying, the probability of $n$ consecutive errors, $P_{e}(n)$, is dependent upon the correlation properties of the channel \cite{zorzi}. The correlation coefficient $\rho$ is
\begin{equation}
    \rho = J_o(2\pi f_d T_s)
\end{equation}
and we define variable
\begin{equation}    
    \phi = \frac{2\gamma_{th}}{1-\rho^2}
\end{equation}
where $J_o$ is the zero-th order Bessel function of first kind and $f_d$ is the Doppler shift.

The probability of error for a single back to back failure is given using the first-order Markov model in \cite{zorzi} as
\begin{eqnarray} \label{eq:ch3.ieee.79}
    P_{bb} = 1 - \frac{Q(\phi,\rho \phi) -  Q(\rho\phi, \phi) }{\exp\left(\gamma_{th}\right) -1}
\end{eqnarray}
where $Q$ is the Marcum Q-function.
Using $P_{bb}$, the probability of $n$ consecutive outages $P_e(n)$ given as 
\begin{equation}\label{eq:ch3:ieee.80}
    P_{e}(n) = P_{bb} \cdot P_{e}(n-1)
\end{equation}
with $P(1)$ given in (\ref{eq:ch3.ieee.71}).

\begin{figure}[t!]
    \centering
   \includegraphics[scale = 0.5]{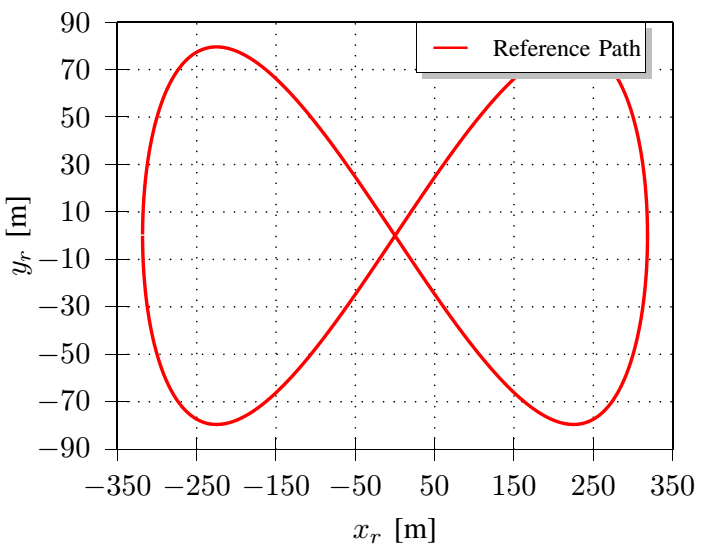}
    \caption{Reference track of an AGV}
    \label{fig:ch3.ieee.10}
\end{figure}



\section{Results} \label{sec:results} 
\subsection{Simulation setup}
\paragraph{Reference path and velocities}
The reference track $X_r(k)$ shown in Figure \ref{fig:ch3.ieee.10} is assumed to be traced by an AGV in $T$ seconds. The AGV starts at position $(x_r(t_0), y_r(t_0)) = (-350,0)$. At time $T$, the AGV should complete the reference path and reach again the initial position, i.e., $X_r(0) = X_r(T)$. The total number of time steps $N_k$ required to reach the final position is given as $N_k = \lceil T/T_s\rceil$. This also determines the reference translational velocity $\nu_r(k)$ and rotational velocity $\omega_r(k)$ of an AGV at every time step. If the time to complete the reference path is lower, higher will be the velocities $\nu_r$ and $\omega_r$. The control input $u(k)$ is sent with $D_i = \unit[78]{bytes}$, on the channel bandwidth $B = \unit[10]{MHz}$, with an average SNR $\gamma_i = \unit[10]{dB}$ to $i-th$ AGV. The bandwidth is equally distributed over $N= \unit[50]{AGVs}$. The values of gain constants $K_x$, $K_y$ and $K_\theta$ in the control law ~\eqref{eq:ch3.ieee.15} are given to be $K_x=\unit[10]{s^{-1}}$, $K_y=\unit[6.4\times 10^{-3}]{m^{-1}}$ and $K_\theta=\unit[0.16]{m^{-1}}$ respectively as in \cite{Kanayama1990}.

\begin{figure}[!ht]
\centering
\includegraphics[scale = 0.45]{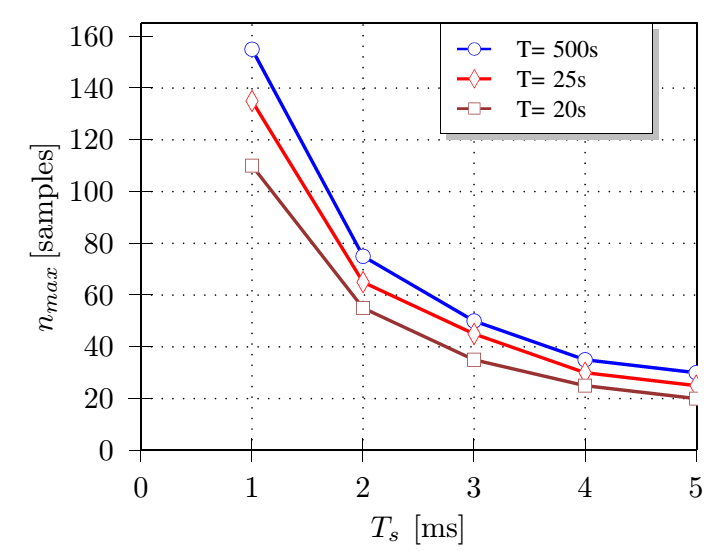}
\caption{Outage tolerance vs. sampling time} 
\label{fig:ch3.ieee.1}
\end{figure}

\begin{figure*}[ht!]
    \centering
     \includegraphics[scale = 0.5]{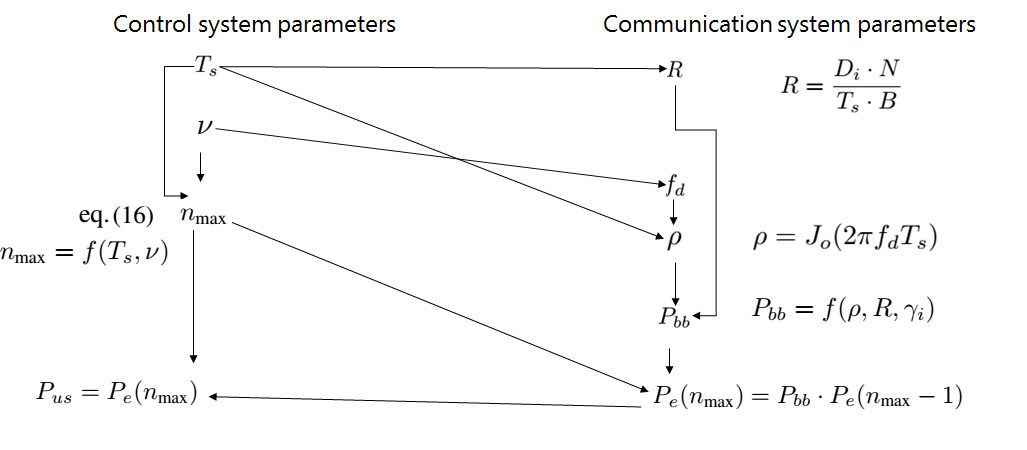}
    \caption{Inter-dependence of control and communication system parameters}
    \label{fig:ch3:2}
\end{figure*}

\subsection{Inter-dependence of control and communication}
In order to study dependencies between the communication and control parameters as shown in Figure \ref{fig:ch3:2}, it is necessary to investigate the relation of parameters within the control domain. 

\paragraph{Outage tolerance and sampling time} \label{sec:delay:SamplingTime}
In this section, the impact of sampling time and an AGV's velocity on the outage tolerance $n_\text{max}$ is presented. The stability of a control system depends predominantly upon the sampling time $T_s$ and $n_\text{max}$. If $T_s$ is very large, the non-linear discrete system will be unstable, as the difference equation in~\eqref{eq:ch3.ieee.34}, becomes non-convergent.  Figure \ref{fig:ch3.ieee.1} shows the outage tolerance as evaluated in \eqref{eq:ch3:ieee:59} for different sampling time $T_s$. Higher the sampling time, lower is the number of tolerable outages $n_\text{max}$ such that the product of outage tolerance and sampling time is almost constant. 

The maximum consecutive outage $n_\text{max}$ is also evaluated for different velocities. The time in which the AGV completes the reference path as shown in Figure \ref{fig:ch3.ieee.10} determines the velocity of an AGV.  Since the complete reference track needs to be traced in less time, the AGV has to drive with higher velocities, and the reference velocities $\nu_r(k)$ and $\omega_r(k)$ have higher rate of change at every time step. Therefore, even a few outages would lead to instabilities at higher velocities, which further implies that the maximum outage tolerance $n_\text{max}$ decreases as the velocity is increased. For $T_s=\unit[1]{ms}$, the outage tolerance is $n_\text{max}=\unit[155]{samples}$ if the total time to trace the reference track is $T=\unit[500]{s}$. The tolerance decreases to $n_\text{max} = \unit[110]{samples}$ for $T= \unit[20]{s}$. It can be followed that if the AGV needs to be driven with high velocity, the channel conditions must be better compared to lower velocities, since control input must reach the AGV with less outages.

\paragraph{Impact of communication on control system} Figure~\ref{fig:ch3:2} shows the inter-dependencies between control and communication domain. The sampling time of the control system influences the spectral efficiency $R$, i\,e., if $T_s$ is higher, lower is the spectral efficiency $R$ for a given bandwidth $B$ and data bits $D_i$. As the channel correlation $\rho$ decreases, the probability of back to back error $P_{bb}$ also decreases. Also, if an AGV is moving with higher velocity $\nu$, the Doppler shift $f_d$ increases and the correlation coefficient $\rho$ decreases. The probability that back-back error occurs $P_{bb}$, as evaluated in \eqref{eq:ch3.ieee.79}, will therefore decrease at higher $f_d$ and $T_s$. If $P_{bb}$ decreases, also the probability of $n$ consecutive errors $P_{e}(n)$ decreases. The maximum tolerable outages $n_\text{max}$ that the AGV control system can withstand also depends upon  $T_s$ and $\nu(k)$ as shown in the previous section. The probability that an AGV goes into an unstable state, $P_{us}$, is the probability that $n_\text{max}$ consecutive errors occur. Therefore, for a given sampling time $T_s$ and translational velocity $\nu(k)$, the probability of $n_\text{max}$ consecutive error is evaluated from ~\eqref{eq:ch3:ieee.80}, i.\,e., $P_{us} = P_{e}(n_\text{max})$.

\begin{figure}[!ht]
\centering
\includegraphics[scale = 0.7]{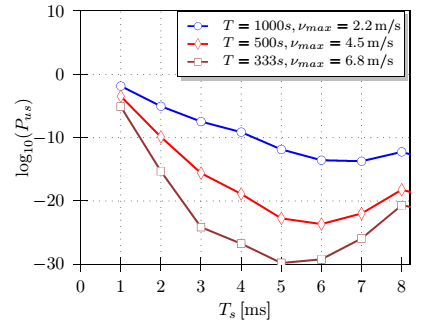}
\caption{Probability  of an unstable AGV over sampling time}
\label{fig:ch3.ieee.5}
\end{figure}

\paragraph{Probability of an unstable system}
Figure \ref{fig:ch3.ieee.5} shows the probability that the control system would experience instability for different values of $T_s$ and $T$. On the one hand, a higher trace time $T$ implies a lower velocity, and therefore a higher channel correlation leading to a higher probability of consecutive errors $P_e(n)$. On the other hand, a lower velocity implies that the AGV can cope with a higher number of consecutive outages as shown in the Figure \ref{fig:ch3.ieee.10}. The question is now which effect is dominating, i\.e., whether the effect of lower channel correlation with increasing velocities outweighs the stronger requirements on $n_\text{max}$.

Figure \ref{fig:ch3.ieee.5} addresses it by showing the probability $P_{us}$ that the system becomes unstable with respect to the sampling time $T_s$ and trace time $T$. For a given sampling time $T_s$, if the AGV velocity is increased (decreasing $T$), the channel correlation decreases, and hence the probability of consecutive error is also lower. On the contrary, at higher velocity the system can tolerate less number of channel outages $n_\text{max}$. Figure~\ref{fig:ch3.ieee.5} shows that at higher velocities the probability of instability decreases. This demonstrates that at higher AGV velocities, the impact of lower consecutive packet errors outweighs the higher outage requirements of the control system. It shows that driving an AGV with higher velocity may indeed reduce the instability probability of a system. Specifically, assuming that if the channel is in a deep fade, a lower AGV velocity will lead to a higher number of consecutive outages (packet losses) due to higher channel correlation. If the consecutive outages exceed  $n_\text{max}$, the system will become unstable. Conversely, driving at higher velocity would reduce the probability of consecutive error, and consequently the probability of an unstable system. 
Furthermore, with a longer sampling time $T_s$, the number of tolerable channel outages $n_\text{max}$ decreases as shown in Figure \ref{fig:ch3.ieee.1}, implying even stronger requirements for a stable control system. Figure~\ref{fig:ch3.ieee.5} shows that the probability of an instability decreases as the sampling time increases. The reason is that increasing $T_s$ decreases the required spectral efficiency $R$ and lowers the channel correlation. The probability of consecutive errors decreases if the channel is less correlated. Hence, for a given velocity, it is better to send the control input less frequently but at higher reliability. Therefore, a higher data rate per AGV is needed which limits the number of admissible AGVs in the system.
It is also shown that at higher velocities the system cannot cope with longer sampling times $T_s$. However, in Figure \ref{fig:ch3.ieee.5}, for $T=$\unit[333]{s}, $P_{us}$ decreases to an optimal sampling point at $T_s = 5$\,s and then increase again. It demonstrates that under strong channel correlation, the control input should be sent at longer time intervals. The probability of instability is higher if the control updates are sent more frequently, due to higher probability of consecutive errors. Moreover, if this optimum $T_s$ is exceeded, the channel conditions do not improve sufficiently to outweigh the stringent requirements on $n_\text{max}$. Additionally, it implies that the channel is uncorrelated and the control updates can be successfully sent. Furthermore, at lower AGV velocity, lower is the Doppler shift. Therefore, the channel is highly correlated even for longer sampling times. For $T=\unit[500]{s}$, i.\,e. the maximum translational velocity $\nu_{max} = \unit[4.5]{m/s}$, the optimal sampling time is \unit[6]{ms}, whereas for $T=\unit[1000]{s}$, the optimal sampling time is \unit[7]{ms}.
This manifests that to reduce the probability of an instability, the control updates should be sent at optimal time interval for a given AGV velocity.


\section{Conclusion}\label{sec:conclusion}
This  paper considered a very specific sandbox example to investigate the interaction of a control system and mobile communication. However, this simple example already revealed the strong inter-dependencies and the potential for joint optimization and design of control and communication systems in industrial automation. In our example, we investigated the impact of an AGV's velocity,  channel outages and sampling time on the stability of a control system. In the future, we will further investigate additional parameters including different re-transmission strategies, the impact of multi-user technologies such as opportunistic scheduling, the maximum number of admissible AGVs in a system under given mobile communication conditions, as well as intelligent control system algorithms which already take into account the characteristics of a mobile communication system.

\section{Acknowledgment}
This research was supported by the German Federal Ministry of Education and Research (BMBF) under grant number
KIS15GTI007 (project TACNET4.0, www.tacnet40.de). The responsibility for this publication lies with
the authors.

\bibliographystyle{IEEEtran}
\bibliography{IEEEabrv,SCC.bib}

\end{document}